\documentclass[11pt]{article}

\usepackage{graphicx}

\begin{document}

\centerline{\LARGE \bf Comment on "Models of Intermediate Spectral Statistics"}

\vspace*{1cm}

\centerline{\bf \Large T. Gorin$^{1}$,
M.~M\"uller$^{2}$ and
P.~Seba$^{3,4}$}

\vspace*{.3cm}

{\it
\begin{center}
$^1$ Centro de Ciencias
F\'{\i}sicas, UNAM, Campus Morelos, Cuernavaca, Mexico,
62251, A.P. 48-3\\

$^2$ Facultad de Ciencias,
UAEM, C.P. 62210, Cuernavaca, Morelos, Mexico\\

$^3$ Institute of Physics, Czech Academy of Science, Cukrovarnicka 10,
Prague, Czech Republic\\

$^4$ Pedagogical University, Department of Physics, Hradec Kralove, Czech Republic\\
\end{center}}

\vspace*{.5cm}

PACS: 03.65, 05.45

\vspace*{1.5cm}
\begin{abstract}

In \cite{Bogom} it was proposed, that the nearest-neighbor
distribution $P(s)$ of the spectrum of the Bohr-Mottelson model is
similar to the semi-Poisson distribution. We show however, that
$P(s)$ of this model differs considerably in many aspects from
semi-Poisson. In addition we give an asymptotic formula for $P(s)$
as $s \to 0$, which gives $P'(0) = \pi\sqrt{3}/2$ for the slope at
$s=0$. This is different not only from the GOE case but also from
the semi-Poisson prediction that leads to $P'(0) = 4$
\end{abstract}
\vspace{1cm}

The motivation for this comment stems from the impression, that the article
``models of intermediate spectral statistics'' can easily be misinterpreted in
two ways: One might be led to believe that (i) the semi-Poisson distribution is
universal, and (ii) the universality class of ``intermediate statistics'' is as
well defined and established as for example the Poisson ensemble or the GOE.
In this comment, we will argue, that both statements are wrong.

The purpose of \cite{Bogom} is to present models which could constitute a
``third'' universality class of systems which show so called ``intermediate
statistics'', previously introduced by Shklovskii in \cite{Shklovskii}.
The Poissonian and the Gaussian ensemble (for definiteness, consider orthogonal
ensembles only) are considered as the first two universality classes in this
list.

As in the Poissonian and in the GOE case, where the respective
members  have common and unique statistical properties, one would
expect the same to hold for the models with intermediate
statistics. In \cite{Bogom} the authors concentrate on the
distribution of nearest neighbor spacings. In the Poissonian case
it is given by $P(s)= \exp(-s)$, in the GOE case it is close to
the well known Wigner surmise $P(s)\approx (\pi/2) \exp(-\pi /4
s^2)$, whereas in the case of the ``intermediate statistics'' the
candidate proposed in \cite{Bogom} is  the semi-Poisson
distribution $P(s)= 4s\; \exp(-2s)$.

In what follows we will show, that the level spacing distribution
in the case of the Bohr-Mottelson model, that represents one of
the candidate systems for the intermediate statistics mentioned in
\cite{Bogom} is in fact very different from the proposed
semi-Poisson distribution. This discrepancy can be   found in fact
already on a figure published in \cite{Montambaux}  but  without
discussing  the problem. However, \cite{Montambaux} gives an overview
over the statistical properties of different variants of the Bohr-Mottelson
model. \\

In \cite{Bogom} the following matrix model (originally introduced by Bohr and
Mottelson) is presented as a possible candidate showing statistical properties 
similar to the semi-Poisson distribution:
\begin{equation}
H_{mn} = e_n \delta_{mn} + t_m t_n \; .
\label{model}
\end{equation}
$H$ is a $N\times N$-matrix,  $e_n$ are mutually independent random
variables uniformly distributed over a finite interval, and the $t_n$ are
chosen with equal absolute value squared $t_n^2 = r$.
The authors of \cite{Bogom}   sketch a procedure for calculating
analytically the 2-point correlation function. For small distances
it should agree with $P(s)$, so that one can derive the slope
$P'(0)= \pi\sqrt{3}/2$ of the spacing distribution at $s=0$. This
slope is different from  the GOE case, where it equals to $\pi^2 /6$
as well as from the slope of the semi Poisson distribution that
equals to $4$.
The difference to the GOE case is remarked, but a similar remark on the 
difference to the semi-Poisson is avoided. An unprejudiced reader might 
believe, that the correlation properties of the matrix model are similar to 
semi-Poisson, even though that the spectral statistics of this model differ
remarkably. \\

In Figure 1 we present the numerical result for $P(s)$ obtained
for an ensemble of 1000 matrices of dimension 750. For the
statistical analysis we used only one third of the states in the
center of the spectral region. The numbers  $e_n$ are uniformly
distributed over an interval $[-1,1]$ and the elements of the
vector $\vec t$ are chosen as $t_i=\sqrt{\alpha/(\pi\rho N)}$,
where $\rho$ is the level density in the center of the spectrum,
$N$ is the dimension of the matrix and $\alpha=10$ is the coupling
constant. (We checked, that  larger coupling does not change the
numerical results). Figure 1.a demonstrates the qualitative
differences in the behavior of $P(s)$ between the random matrix
model and the semi-Poisson distribution. The slope at $s=0$ is
smaller, the maximum of $P(s)$ is slightly shifted to the right,
and for values $s>1$ it shows significant deviations well above of
the statistical error. Hence the spacing distribution of the
Bohr-Mottelson model is not close to semi-Poison.
\begin{figure}
\begin{center}
\includegraphics[scale=0.7]{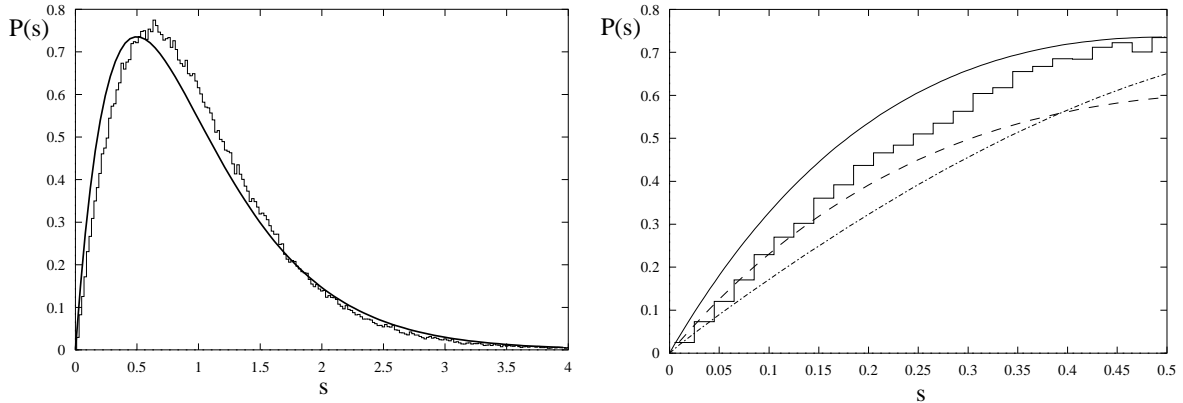}
\caption{(a) $P(s)$ for the model (\ref{model}) compared with the the semi-Poisson distribution,
(b) the same as in (a) for short distances. In addition the theoretical result
(\ref{3point}) is drawn as a dashed line
and the GOE result as a dashed dotted curve.}
\end{center}
\end{figure}

Figure 1.b. shows a magnification of the interval $0 \le s \le 1/2$ using
the same data as in figure 1.a. In this figure we additionally plotted
the asymptotic result (\ref{model}) for the present model as a dashed line.
The basic idea of the approximation is the following:
In order to get a short distance between two neighbored levels in the
spectrum of $H$, three eigenvalues of $H_0$ have to come close together.
Then the levels which are farther away can be neglected. Therefore we can
restrict the sum
\begin{equation}
K(E) = \sum_{i=1}^N \frac{t_i^2}{E-e_i}
\end{equation}
whose roots define the eigenvalues of $H$, to those terms with the three
consecutive eigenvalues. Resolving for the two roots, calculating their
distance, and averaging over the levels $e_i$ leads to the
following formula
\begin{equation}
P(s)= \frac{9 s}{4} \intop_0^{\pi/2} d\phi \frac{{\rm exp} \left[
-\frac{3s}{2} (\cos\phi + \sin\phi)/\sqrt{1+\sin(2\phi)/2} \right ]}
{1 + \sin(2\phi)/2} \; .
\label{3point}
\end{equation}
The dashed curve in Figure 1.b. is obtained from a numerical integration
of (\ref{3point}). At short distances, this approximation describes the
numerical data much better than the semi-Poisson. A Taylor expansion of the
integrand of (\ref{3point}) gives $P'(0)= \pi\sqrt{3}/2$ for the slope at
$s=0$, the same result as found in \cite{Bogom}.

A detailed numerical investigation of several statistical properties of the
type of models can be found also in \cite{dittes, gorin, Montambaux}.\\

To conclude, if one wants to insist on the introduction of a ``third''
universal ensemble, one should possibly use a criterium similar to the 
following (cited from \cite{Bogom}: ``\ldots the main features are (i) the 
existence of level repulsion (as in random matrix ensembles), and (ii) slow 
(approximately exponential) fall-off \ldots ''. Even though this definition is 
quite ``spongy'', it seems to be the only way to make sure, that the systems 
discussed fit into this class.

\textsc{Acknowledgements} M. M\"uller acknowledges financial support
from the CONNACyT (No.32101-E).

\end{document}